\documentclass[a4paper, preprint, jctc]{revtex4}
\usepackage{amsmath}
\usepackage{amsfonts}
\usepackage{graphicx}
\usepackage{color}
\usepackage{lscape}
\usepackage{amssymb}
\usepackage{threeparttable}
\usepackage{multirow}
\usepackage{framed}

\begin{document}

\title{Assessment of various natural orbitals as the basis of large active space density-matrix renormalization group calculations}
\author{Yingjin Ma}
\author{Haibo Ma}
\thanks{Author to whom correspondence should be addressed. Electronic address: haibo@nju.edu.cn}
\affiliation{Key Laboratory of Mesoscopic Chemistry of MOE, School of Chemistry and Chemical Engineering, Institute of Theoretical and Computational Chemistry, Nanjing University, Nanjing 210093, China}

\date{Latest revised on \today}

\begin{abstract}

It is well-known that not only the orbital ordering but also the choice of the orbitals themselves as the basis may significantly influence the computational efficiency of density-matrix renormalization group (DMRG) calculations.
In this study, for assessing the efficiency of using various natural orbitals (NOs) as the DMRG basis, we performed benchmark DMRG calculations with different bases, which included the NOs obtained by various traditional electron correlation methods, as well as NOs acquired from preliminary moderate DMRG calculations (e.g., preserved states less than 500). The tested systems included N$_2$, transition metal Cr$_2$ systems, as well as 1-D hydrogen polyradical chain systems under equilibrium and dissociation conditions and 2-D hydrogen aggregates. The results indicate that a good compromise between the requirement for low computational costs of acquiring NOs and the demand for high efficiency of NOs as the basis of DMRG calculations may be very dependent on the studied systems' diverse electron correlation characteristics and the size of the active space.
It is also shown that a DMRG-complete active space configuration interaction (DMRG-CASCI) calculation in a basis of carefully chosen NOs can provide a less expensive alternative to the standard DMRG-CASSCF calculation and avoid the convergence difficulties of orbital optimization for large active spaces. The effect of different NO ordering schemes on DMRG-CASCI calculations is also discussed.

\end{abstract}

\maketitle

\section{Introduction}

In recent years, novel multi-reference approaches based on the density-matrix renormalization group (DMRG) algorithm \cite{dmrg1, dmrg2} have been introduced into the theoretical chemical community for static quantum chemical calculations \cite{poly-dmrg, ab-dmrg, Daul00, Chan02, ab-dmrg2, Chan03, Chan04, dmrg-casscf-diis, ab-dmrg4, Barcza11, Boguslawski11, Boguslawski12, Boguslawski12-2, local-fano, Hachmann06, Hachmann07, bi-dmrg, Zgid08, dmrg-rdm, dmrg-casscf, dmrg-scf, dmrg-pt2, dmrg-ct, dmrg-opt, Legeza03-0, Legeza03, Legeza03-1, Legeza04, Moritz05, Moritz06, Rissler06, Mitrushenkov01, Mitrushenkov03, dmrg-qc-rev, Marti08} and real-time electronic wavefunction evolution \cite{real-time, real-time1, real-time2}. DMRG, first developed by White in 1992, uses the eigenvalues of the subsystem's reduced density matrix as the decimation criterion of the Hilbert space, and is regarded as an efficient alternative to full configuration interaction (FCI) in quantum chemical descriptions. The computational cost of the DMRG algorithm is largely caused by the scaling of \emph{O(k$^3$M$^3$)} + \emph{O(k$^4$M$^2$)}, where $k$ is the number of active space orbitals, and $M$ is the number of renormalized states, which usually determines the accuracy of the method \cite{Chan02, ab-dmrg4}. This cost is far less than that of the FCI, and usually the DMRG method can deal with dozens of electrons in dozens of orbitals \cite{dmrg-qc-rev}.
Considering DMRG's significant advantage of computational efficiency, quantum chemists have successively combined it with various \emph{ab initio} methodologies, such as FCI \cite{ab-dmrg, Daul00, Chan02, Mitrushenkov01, Mitrushenkov03, Chan03, Chan04, Rissler06, Legeza03-0, Legeza03, Legeza03-1, Legeza04}, complete active space configuration interaction (CASCI) \cite{Moritz05, Moritz06, bi-dmrg, local-fano, Barcza11, Boguslawski11, Boguslawski12, Boguslawski12-2, Hachmann06, Hachmann07, Marti08, Zgid08, dmrg-rdm, ab-dmrg2, ab-dmrg4} and complete active space self-consistent field (CASSCF) \cite{dmrg-casscf, dmrg-scf, dmrg-casscf-diis} as well as complete active space perturbation theory (CASPT2) \cite{dmrg-pt2}, and canonical transformation theory (CT) \cite{dmrg-ct}.

DMRG has an innate advantage for the descriptions of one-dimensional (1-D) systems, and when localized molecular orbitals (LMOs) are used, the active space can be extended to about 100 orbitals with very good accuracy \cite{Hachmann06}. However, there are still difficulties in the localization process for large systems when large basis sets including polarized and diffused functions are used \cite{loc-difficult1, loc-difficult2}. On the other hand, more importantly, not all of the studied systems belong to the 1-D framework, and as such, orbital localization might be inefficient.
For general systems, the HF canonical molecular orbitals (CMOs) are often used as the basis set for DMRG-based calculations. If necessary, a further orbital optimization procedure \cite{dmrg-casscf, dmrg-scf} is implemented to get more reasonable orbitals or the orbital reordering procedure by quantum information theory \cite{Legeza03, Moritz05, Rissler06} is performed to reduce the interaction ranges in the DMRG ``sweep" process. Nevertheless, orbital optimization procedures used in the DMRG calculations mainly inherit the traditional multi-configuration self-consistent-field (MCSCF) or CASSCF algorithm, wherein the orbital optimization procedure involve many ``macro-iterations" and often encounters convergence difficulties for large active spaces. In DMRG-CASSCF, one macro-iteration is defined as one DMRG-CASCI calculation followed by orbital optimization and integral transformation (involving about $\frac{3}8N^4M+3N^3M^2$ operations, where $N$ and $M$ represent the total orbitals and internal orbitals, respectively) \cite{MCSCF2} or a transformation of Coulomb and exchange operators (about $\frac{3}2N^2M^3$ operations) \cite{MCSCF2}, and the precess is usually very computationally expensive.

On the other hand, many DMRG-CASSCF calculations use CMOs from Hartree-Fock (HF) calculations as the initial guess, but the virtual orbitals in CMOs have little or no resemblance with correlated orbitals. Therefore, the starting point with CMOs is arbitrary, and convergence may be obtained to an undesired stationary point \cite{caspt2no}.
In addition to CMOs and LMOs, natural orbitals (NOs), have also been introduced as the basis of DMRG-FCI calculations \cite{Mitrushenkov01, Mitrushenkov03, Rissler06}, as well as later for DMRG-CASCI and DMRG-CASSCF calculations \cite{Marti08, dmrg-scf, Boguslawski11, ab-dmrg4, Boguslawski12, Boguslawski12-2}.

NOs, firstly introduced in 1955 by L\"{o}wdin \cite{NO}, can be obtained from the one-electron reduced density matrix (1-RDM) and sorted by natural orbital occupation numbers (NOONs).
Numerical computations \cite{INTRO} show that the electron correlation effect is mainly affected by the NOs with large NOONs,
,and as such, discarding the NOs with small NOONs could be an efficient approach to reduce the computational burden with marginal loss of accuracy.
Beyond their usage in truncating virtual orbitals, the adoption of NOs also has important advantages in the multi-reference quantum chemistry for choosing the active space and generating starting orbitals for orbital optimization iterations.
These aspects are vital in DMRG-based calculations, because the systems studied by DMRG are usually much larger and more complicated than normal systems, which can be treated by traditional multi-reference methods. A reasonable active space and an appropriate set of starting orbitals are obviously highly desired for DMRG-based calculations.
However, choosing appropriate orbitals and a reasonable active space is no ``black box" task at this stage.
When using NOs, one can simply define an appropriate threshold for the NOONs to select the active space, say 0.02-1.98 \cite{INTRO}. Another more rigorous criterion has recently been introduced by Landau et al. \cite{noselect} using the ratio between the cumulative occupation number and the total number of electrons, and this algorithm tends to be less sensitive to the studied systems and basis sets \cite{exp-landau}.
Since the NOs are also self-optimized through the electron correlation calculation within a large or full orbital space, they can be used as a computationally cheap alternative to the optimized orbitals obtained by CASSCF calculations \cite{UNO-casci, NO-casci, IVO, NO-ddci}. For example, using CAS(2,2) with a difference-dedicated configuration interaction (DDCI) restricted to the valence $\pi$ orbitals in organic magnetic systems, one can obtain the NOs of the same quality as the CAS(full valence $\pi$)SCF orbitals when determining magnetic couplings \cite{NO-ddci}. Actually, sometimes NOs might be even more suitable than CASSCF-optimized orbitals for further higher-level electron correlation calculations because they inherently contain dynamic correlation information, provided they are obtained from the electron correlation calculations conducted within a large active space.

In principle, NOs from high-level quantum chemical calculations can provide a more reasonable basis for DMRG-based calculations, because their electron correlation information is already very accurate. Nevertheless, such high-level preliminary calculations will be obviously very expensive and sometimes unaffordable. Therefore, a good compromise between the requirement for low computational costs of acquiring NOs and the demand for high efficiency of NOs as the basis of DMRG calculations would greatly benefit practical studies. Although NOs have been widely used as the basis of DMRG-based calculations in the past decade \cite{Mitrushenkov01, Mitrushenkov03, Rissler06, Marti08, dmrg-scf, Boguslawski11, ab-dmrg4, Boguslawski12, Boguslawski12-2}, detailed benchmark evaluations of the performances of various NOs and other types of molecular orbitals in DMRG-based calculations are still absent, and the determination of the types of NOs that can provide the most reasonable basis for DMRG calculations is still unknown.

In this study, we assess the efficiency of various NOs used in large active space DMRG calculations through benchmark tests of DMRG-CASCI calculations using NOs generated by unrestricted Hartree-Fock (UHF), configuration-interaction singles and doubles (CISD), second-order M\o ller-Plesset perturbation theory (MP2), CASPT2 and multi-reference configuration interaction (MRCI). We also list the results from HF CMOs, Kohn-Sham MOs, as well as the CASSCF-optimized MOs.
Nevertheless, the NOs generated by CASPT2 may be limited by the size of the complete active space; thus, a series of results using NOs generated by tentative DMRG-CI calculation with moderate preserved states are also given. With moderate preserved states (e.g., $M\leq$ 500), the number of active orbitals in feasible DMRG calculations can increase to about 100. DMRG-CASSCF calculations are also implemented, and detailed comparisons between results of DMRG-CASCI and DMRG-CASSCF are carefully performed.

The outline of this study is as follows. In Sec. II, computational details are introduced; in Sec. III, demonstrative computations of N$_2$, Cr$_2$, as well as 1-D and 2-D hydrogen aggregates are presented under equilibrium and bond dissociation conditions, and finally, our results are summarized and concluded in Sec. IV.

\section{Computational details}

The DMRG-CASCI calculations and preliminary DMRG-CI calculations were implemented by the Block code \cite{ab-dmrg4} developed by Chan's group. The NOs for DMRG-CASCI calculations were selected according to their NOONs and grouped by $A_g$, $B_{3u}$, $B_{2u}$, $B_{1g}$, $B_{1u}$, $B_{2g}$, $B_{3g}$, and $A_u$. If no other state, in the DMRG calculation, the orbital groups were ordered by $A_g$, $B_{1u}$, $B_{3u}$, $B_{2g}$, $B_{2u}$, $B_{3g}$, $B_{1g}$, and $A_u$; and in each group the orbitals were aligned in descending order by their NOONs. The orbital optimization procedures in DMRG-CASSCF calculations were implemented by our own code, which is mainly based on the method of Werner and Meyer (Ref.\cite{MCSCF1}), which contains not only the first- and second-order terms but also part of the higher-order terms in the independent orbital parameters \cite{mc-nr}.
The integrals used in orbital optimization were extracted from GAMESS \cite{gamess}.
The number of preserved states used in DMRG-CASCI or DMRG-CASSCF calculations is $M$=2000 unless otherwise stated. The NOs from UHF, MP2, CISD, MRCI and CASPT2, as well as the KS MOs, were obtained from Molpro \cite{molpro}, and the NO's integrals used in the Block code were also obtained from Molpro. However, multi-configuration perturbation theory may encounter ``intruder" problems even in the ground-state calculation due to the size of the active space in the Cr$_2$ system \cite{caspt2-ls}, and therefore, the level shift parameter \cite{caspt2-ls} is necessary to correct the zeroth-order Hamiltonian. Here the \emph{g1} zeroth-order Hamiltonian \cite{caspt2-g1} and a level shift parameter 0.30 were used when applying the CASPT2 methods to obtain the NOs.

Herein we also exploited the DMRG-CI algorithm to generate sets of NOs for further higher-level DMRG-CASCI calculations.
First, we use the DMRG-CI to do a large active space calculations with gradually increasing $M$ values. Second, get the natural orbitals $C'$ according to the transform matrix U which is obtained from the DMRG-CI's 1-RDM calculations ($C'=CU$,  where $C$ usually represents the canonical molecular orbitals). Third, see whether the $M$ values were sufficient by checking the dependence of the NOONs on the $M$ values adopted in the DMRG-CI calculations. Finally, if the changes in the NOONs were tiny or converged, one can refer to the NOONs to determine the active orbitals, which were then taken as the basis for the subsequent DMRG-CASCI calculation.

\section{Results and discussion}

\subsection{N$_2$}

\begin{figure}[!htp]
\centering
\includegraphics[scale=0.5]{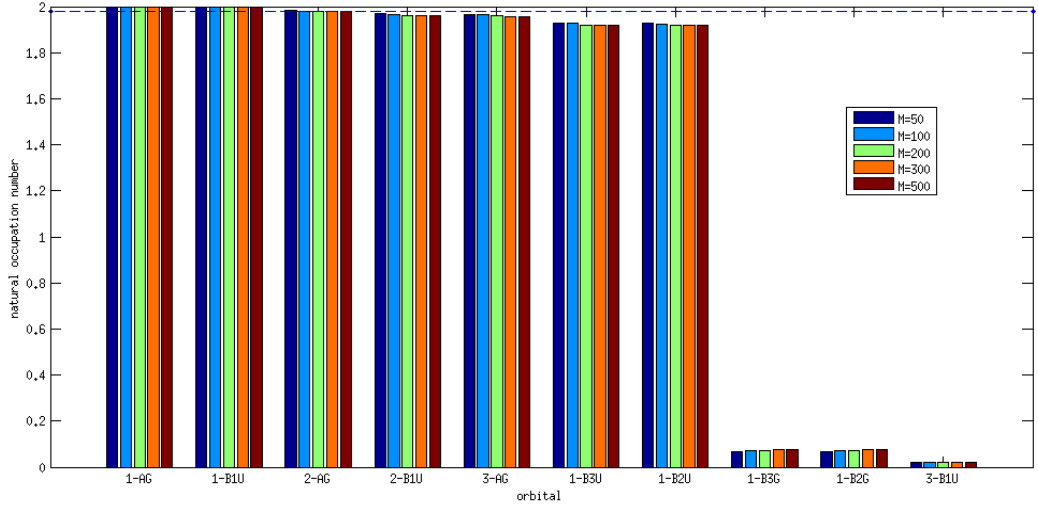}
\caption{The NOONs in the 1.0Re spacing N$_2$ systems (equilibrium) obtained by DMRG-CI calculations.}
\label{fig-no1}
\end{figure}

\begin{figure}[!htp]
\centering
\includegraphics[scale=0.5]{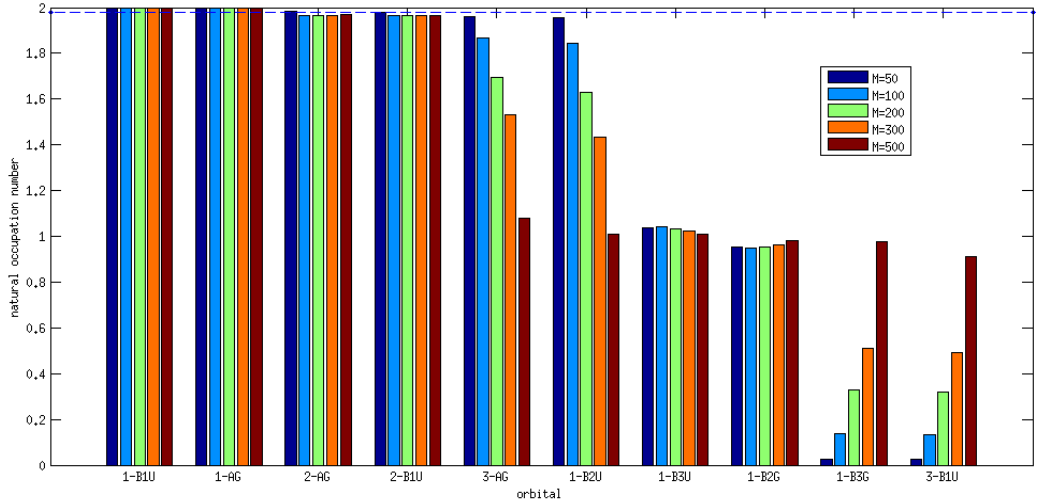}
\caption{The NOONs in the 3.0Re spacing N$_2$ systems (dissociative) obtained by DMRG-CI calculations.}
\label{fig-no2}
\end{figure}

A simple N${_2}$ system is taken as our first example. As a main group element, nitrogen's active space is easily defined, only choosing the valence orbitals into the active space is ideal in most cases \cite{INTRO}, because the energy gap between 2p and 3s orbitals is relatively large. Here we choose two different inter-molecular spacings (1.0$R_e$ and 3.0$R_e$, for $R_e$=1.0976\AA), which represented equilibrium and dissociation conditions, respectively. The basis set used here was Dunning's correlation consistent basis sets cc-pvtz with 60 molecular orbitals assembled from 70 atomic bases in spherical coordinates. Three different active spaces were used in our DMRG-CASCI calculations: the first space (Space-1) is 14 electrons in the valence orbitals and 1s orbitals; the second space (Space-2) inherits the first space and added 8 more orbitals, and admitted or rejected NOs based upon their NOONs, or added 3s and 3p orbitals in the HF or KS MOs case; the third space (Space-3) follows Space-2 and 10 more orbitals were added, which also based on NOONs or simply chose 3d orbitals in HF or KS MOs cases.
The DMRG-CI's NOs were derived from preliminary DMRG-CI calculations with gradually increasing $M$ values of the full 60 MOs. A part of the NOONs obtained using different $M$ values is illustrated in Fig.~\ref{fig-no1} and Fig.~\ref{fig-no2}. In the equilibrium configuration, as shown in Fig.~\ref{fig-no1}, the NOONs obtained by different DMRG-CI calculations with different $M$ values are nearly uniform. The results suggest that the Hartree-Fock approximation should be a good starting point for a description of this configuration, and later, the consideration of only the dynamic correlations could already give a very good correlated picture for N$_2$. When considering the dissociation condition, the NOONs obtained by DMRG-CI calculations with different $M$ values remarkably varied,
since the non-dynamic correlations needed in this condition can not be acquired by the insufficient number of DMRG's preserved states.

After obtaining all of the NOs, the DMRG-CASCI calculations were then implemented on the basis of the described orbital spaces. The ground-state energy results are listed in the Table.~\ref{tab-N2}. From the table, in the equilibrium condition, except for the HF CMOs, UHF NOs, Kohn-Sham MOs, and CASSCF(6,6) optimized MOs, all the other electronic correlations methods can give good sets of NOs ---- The deviations between the results of DMRG-CASCI and DMRG-CASSCFare usually only about 1 or 2 mH, and these deviations are stable after enlarging the active spaces.
The results obtained suggest that these NOs are reasonable and are already very similar to the final optimized orbitals obtained by CASSCF or DMRG-CASSCF calculations.
The NOs from preliminary DMRG-CI calculations give a set of similar results: in Space-1 (14e,10o), all of these DMRG-CI's NOONs support that the NOs derived from 1s, 2s and 2p atomic orbitals play most important roles, and the following DMRG-CASCI results are also nearly identical. In the larger active space (Space-2 with 14e and 18o), a slightly larger deviation (about 4 mH) emerges between the $M\leq$100 series and $M\geq$200 series, and some DMRG-CASCI results are below those of DMRG-CASSCF, which are both caused by the different choices of active orbitals using NOONs, as indicate in the footnote of Table.~\ref{tab-N2}. However, deviations caused by the different choice of orbitals become very small when the active space is enlarged. In Space-3 (14e,28o), although there are still diversities in choosing orbitals among different DMRG' NOs and MP2, CISD etc., the deviations among them are only 1 mH, since the NOONs in the higher NOs will be very small, which indicates that those orbitals have marginal contributions to the correlated energy, and such, the energy changes caused from different sets of orbitals should be also inconspicuous. We can also observe that the DMRG-CASCI calculations with UHF NOs or CASSCF MOs display worse results in Space-2(14e,18o) and Space-3(14e,28o) than in Space-1(14e,10o) due to the lack of dynamic correlations between these two types of orbitals.

\begin{table}[!hbp]
 \centering\footnotesize
 \begin{threeparttable}
 \caption{\label{tab-N2}Calculated CASSCF/DMRG-CASSCF ground state energies (in a.u.) and deviations (in a.u.) and their deviations from DMRG-CASCI values in the N$_2$ system. The cc-pvtz basis and D2h symmetry was used here.}
  \begin{tabular}{lcccccccccccccc}
  \hline
  \hline
   &    &&  \multicolumn{3}{c}{1.0Re spacing} &&  \multicolumn{3}{c}{3.0Re spacing} \\
   \cline{4-6}  \cline{8-10}
   & Orbital type && {(14e,10o)$^a$} & {(14e,18o)$^b$} & (14e,28o)$^c$  & & (14e,10o)$^a$ & (14e,18o)$^d$ & (14e,28o)$^e$ \\  
   \hline
   & DMRG-CI(14e,60o) NOs \\
   & $M$=50  && 0.00153 & 0.00122 & 0.00140 && 0.23557 & 0.21588 & 0.03819\\
   & $M$=100 && 0.00130 & 0.00142 & 0.00153 && 0.02157 & 0.02293 & 0.02743\\
   & $M$=200 && 0.00163 & -0.00282 & 0.00148 && 0.01205 & 0.01545 & 0.01490\\
   & $M$=300 && 0.00182 & -0.00237 & 0.00170 && 0.01018 & 0.01262 & 0.01032\\
   & $M$=500 && 0.00216 & -0.00224 & 0.00183 && 0.00027 & 0.00226 & 0.00081\\
   \hline
   & HF CMOs       && 0.08169  & 0.08761 & 0.07978 && 0.05142 & 0.02291 & 0.00598 \\
   & Kohn-Sham MOs && 0.07078  & 0.08135 & 0.07733 && 0.01646 & 0.02122 & 0.05908 \\
   & NOs (UHF)$^f$ && 0.02101  & 0.09059 & 0.13278 && 0.00031 & 0.03474 & 0.12754 \\
   & NOs (MP2)     && 0.00262  & 0.00157 & 0.00256 &&    -     &    -    &    -    \\
   & NOs (CISD)    && 0.00161  & 0.00131 & 0.00199 && 0.03208 & 0.02314 & 0.00686 \\
   & MOs (CASSCF)$^g$ && 0.00732 & 0.07046 & 0.07794 && 0.00388 & 0.02308  & 0.06269 \\
   & NOs (CASPT2)$^g$ && 0.00072 & 0.00184 & 0.00155 && 0.00077 & 0.00242 & 0.00173 \\

  \hline
 & CASSCF/DMRG-CASSCF && -109.13201 & -109.25577 & -109.33890 && -108.79489 & -108.87457 & -108.99901 \\
  \hline
  \hline
  \end{tabular}
 \begin{tablenotes}
\item[a] Orbitals are (3,1,1,0,3,1,1,0) with irreducible representation order $A_g$, $B_{3u}$, $B_{2u}$, $B_{1g}$, $B_{1u}$, $B_{2g}$, $B_{3g}$, $A_{u}$.
 \item[b] (5,2,2,0,5,2,2,0) in HF, KS and CASSCF MOs; (6,3,3,0,4,1,1,0) in MP2, CISD, CASPT2 and DMRG-CI($M=50$, $M=100$) NOs as well as the DMRG-CASSCF calculation; (6,3,2,1,4,1,1,0) in other DMRG-CI NOs.
 \item[c] (7,3,3,1,7,3,3,1) in HF, KS, CASSCF MOs and CASPT2 NOs; (7,3,3,2,6,3,3,1) in CISD and MP2 NOs; (7,4,3,1,6,3,3,1) in DMRG-CI NOs and DMRG-CASSCF.
 \item[d] (5,2,2,0,5,2,2,0) in all cases.  \quad $^e$ (7,3,3,1,7,3,3,1) in all cases.
 \item[f] No symmetry was used here, three pairs of orbitals were mixed (HOMO - LUMO, HOMO-1 - LUMO+1 and HOMO-2 - LUMO+2). The S$^2$ values were 0.00000258(1.0Re) and 3.00907136(3.0Re).
 \item[g] (6e,6o) as the active space
 \end{tablenotes}
 \end{threeparttable}
\end{table}

In the longer-separation case in which the multi-reference character is strong, the situation may be a bit more complicated. The MP2 calculation can hardly converge under such a condition. Other methods can converge, but only part of them can provide reasonable NOs. In Space-1(14e,10o), the NOs from single-reference methods such as HF and CISD give very poor results, and dozens of $mH$ deviations emerge as compared to the CASSCF results. The KS orbitals yield better results, but remains a deviation of about 16 $mH$ remains. Only the UHF and the multi-reference CASPT2 methods provide reasonable NOs, the deviations being about only 1 $mH$ when one performs a CAS-CI calculation. For the NOs derived from DMRG-CI calculations, NOs with \emph{M} = 50 give the worst results with a deviation of about 0.2 hartree, since with the limited number of preserved states, DMRG can only recover a small part of the correlations, which is far from sufficient. The results can be improved by gradually increasing the number of preserved states: with $M$ = 100, the deviation decreases to 21 $mH$ and in the $M$ = 500 case, the deviation is reduced to less than 1 $mH$. Such progressive behavior matches well with the NOONs information shown in Fig.~\ref{fig-no2}. When enlarging the active space, only the NOs derived from CASPT2 or DMRG-CI with $M$ = 500 are reasonable, because of the strong multi-reference picture under the dissociation condition. Nevertheless, it can be reasonably suggested that the results from CISD's NOs will gradually approach to those from CASPT2' NOs along with a gradually enlarged active space. The reason is that the static correlation can not be well described at CISD level, but the DMRG-CASCI calculations with an enlarged active space will gradually recover the static correlations while only a small part of the dynamic correlations remain in the out space. Meanwhile, these dynamic correlations could be partly recovered with NOs derived from CISD calculations. These two aspects indicate the NOs from CISD could be comparable to those from CASPT2 if the active space used is sufficiently large.

\subsection{Cr$_2$}

Next, we consider the transition metal chromium dimer (Cr$_2$) systems. The basis set used here was also Dunning's correlation consistent basis set cc-pvtz, with 136 molecular orbitals assembled from 168 atomic base in spherical coordinates. The core orbitals (1s, 2s and 2p) were all kept frozen in all calculations. As a transition metal, Cr typically has partially filled d orbitals and its electronic structure is significantly more complicated than that of main group elements, and as such, the multi-reference methods such as CASPT2 are usually necessary. Besides the NOs from CASPT2, those from single-reference methods were also used, such as CISD. Here we also used the NOs from preliminary DMRG-CI calculation in the full orbital space except core orbitals (1s, 2s and 2p).
Three different active spaces were chosen for the DMRG-CASCI calculations of Cr$_2$. The first active space (Space-1) consisted of 12 electrons in 12 orbitals, including the 3d and 4s orbitals, which is also a popular choice in previous multi-reference calculations. The second active space (Space-2 with 12e and 28o) inherits the first  and added 16 more orbitals, and admitted or rejected NOs based on their NOONs, or added the orbitals according to the orbital energies in the HF or KS MOs case (mainly 4p and 4d). The third space (Space-3 with 12e and 42o) follows Space-2 and adds 14 more orbitals that also based on NOONs or according to orbital energies in HF or KS MO cases.

The ground-state energy results are listed in Table.~\ref{tab-Cr2}. In the near-equilibrium (r=1.5\AA) condition, the behavior of the NOs in the Cr$_2$ system is similar to that of the NOs in the former N$_2$ system, except that the MP2 calculation is not available because of the multi-reference character of the Cr$_2$ system. The DMRG-CASCI calculations from HF CMOs and KS MOs give very poor results, the deviations between these results and those from CASSCF or DMRG-CASSCF being about 0.2 hartree. Orbitals from CASSCF provide the best result of all those in Space-1(12e,12o), since it serves as a reference, while large deviations exist in Space-2(12e,28o) (about 0.12 hartree) and Space-3(12e,42o) (about 0.15 hartree) due to the lack of dynamic correlations. Here the NOs from CISD and CASPT2 provide good basis sets for the DMRG-CASCI calculation. In Space-1(12e,12o), the deviation of the result based on CISD's NOs is about 13 $mH$, while the deviation of that based on CASPT2's NOs is only about 2 $mH$, since CASPT2 describes both the non-dynamic and dynamic correlation well, but CISD can only describe the dynamic correlation well. Upon enlarging the active space, the NOs from CISD provide results similar to those from CASPT2 in Space-2(12e,28o) and Space-3(12e,42o), since the non-dynamic correlation is well described by the DMRG-CASCI calculation conducted in a large active space, and at the same time, the dynamic correlations in the out-orbitals are recovered by CISD or CASPT2 calculations. Here we also list DMRG-CASCI results based on the NOs from the preliminary DMRG-CI calculation. It can be observed that the results based on NOs from DMRG-CI are comparable to those from CISD, which indicates that the dynamic correlations can also be recovered from the moderate DMRG-CI calculation. However, one may notice that under near-equilibrium (r=1.5\AA) condition DMRG-CASCI calculated ground state energy doesn't decrease monotonically with the increasing $M$ values used in preliminary DMRG-CI(12e,118o) calculations.
In fact, the eigenvalue (ground state energy) in our DMRG-CI(12e,118o) calculation is always descending with the increasing $M$,
however, the eigenvectors usually have much slower convergence behavior
than the eigenvalue. As such, there are no guarantee that increasing insufficient $M$ values from 100 to 300 can produce optimal 1-RDM and NO basis for other DMRG-CASCI calculations with different active spaces ((12e,12o) or (12e,28o) or (12e,42o)) and consequently they will not certainly bring in ideal monotonically decreasing ground state energy behavior for these DMRG-CASCI calculations within other active spaces.

 \begin{table}[!hbp]
 \centering\footnotesize
 \begin{threeparttable}
 \caption{\label{tab-Cr2}Calculated ground state energies (in a.u.) using the various starting orbitals and active spaces in the Cr$_2$ system with cc-pvtz basis sets and the D$_{2h}$ symmetry. Here the Hartree-Fock energies of -2085.99908 a.u. (in 1.5\AA) and -2085.35097 a.u. (in 2.8\AA) are taken as references.}
  \begin{tabular}{lcccccccccccccc}
  \hline
  \hline
 &                                 &&& \multicolumn{3}{c}{1.5\AA}  && \multicolumn{3}{c}{2.8\AA}  \\
 \cline{5-7} \cline{9-11}
  & \multicolumn{2}{c}{Orbital type} && {(12e,12o)$^{a,b}$} & (12e,28o)$^{a}$ & (12e,42o)$^{a}$ &&  {(12e,12o)$^{a,b}$} & (12e,28o)$^a$ & (12e,42o)$^a$\\
  \hline
 & DMRG-CI(12e,118o) NOs &&  \\
 & $M$=100 &&& -0.62077 & -0.76943$^c$ & -0.83536$^e$ &&-0.95173 & -1.40102$^g$ & -1.45028$^i$\\
 & $M$=200 &&& -0.61267 & -0.77312$^c$ & -0.83830$^e$ &&-0.95173 & -1.40599$^g$ & -1.45615$^i$\\
 & $M$=300 &&& -0.61103 & -0.76975$^c$ & -0.83823$^e$ &&-1.03297 & -1.41056$^g$ & -1.46400$^i$\\
 \hline
 & HF CMOs       &&& -0.21274  & -0.53392$^d$ & -0.65446$^f$ &&-1.14666 & -1.30573$^h$ & -1.37491$^j$  \\
 & Kohn-Sham MOs &&& -0.27755  & -0.58063$^d$ & -0.66467$^f$ &&-1.01297 & -1.31505$^h$ & -1.38455$^j$\\
 & NOs (MP2)     &&&      -       &      -      &      -   &&      -      &      -      &      -         \\
 & NOs (CISD)    &&& -0.61168  & -0.77733$^d$ & -0.84146$^f$ &&-1.14815 & -1.43312$^h$ & -1.50016$^j$\\
 & MOs (CASSCF$^o$) &&& -0.62410  & -0.65486$^d$ & -0.69245$^f$ &&-1.33799 & -1.35718$^h$ & -1.38423$^j$\\
 & NOs (CASPT2$^o$) &&& -0.62236  & -0.77862$^d$ & -0.83859$^f$ &&-1.33796 & -1.46714$^h$ & -1.52019$^j$\\
  \hline
& CASSCF/DMRG-CASSCF  &&& -0.62410 & -0.77731$^k$ & -0.84216$^l$ && -1.33799 & -1.45717$^{m,uc}$ & -1.51295$^{n,uc}$ \\
  \hline
  \hline
  \end{tabular}
 \begin{tablenotes}
 \item[a-o] Please see the appendix
 \item[uc] This results is not converged.
 \end{tablenotes}
 \end{threeparttable}
\end{table}

Under the dissociation condition (r=2.8 \AA) the electronic structure becomes considerably more complicated because of the stronger multi-reference electronic picture. Here our DMRG-CASSCF calculations converge with great difficult in Space-2(12e,28o) and Space-3(12e,42o) owing to very complicated electronic structures. It can be seen from the table.~\ref{tab-Cr2} that only the calculation based on the NOs from CASPT2 give results similar to those of CASSCF in Space-1, and these results are also the best results obtained in Space-2 and Space-3. The results based on the MOs from HF and KS calculations are still not acceptable. The results from CISD's NOs exhibit a large deviation (about 0.2 hartree) compared to the CASSCF results in Space-1, but such deviations reduced when enlarging the DMRG-CASCI active space, such that deviation (according to the DMRG-CASCI results from CASPT2's NOs) reduced to 30 $mH$ in Space-2 and 20 $mH$ in Space-3. This behavior is similar with the N$_2$ system under the dissociation condition. The NOs derived from the preliminary DMRG-CI calculation may lose efficiency, and corresponding DMRG-CASCI results are difficult to compare with those from CASPT2's NOs. This fact implies that the NOs from the preliminary DMRG-CI calculations may be not suitable for such complicated systems without access to sufficient preserved states. So, the DMRG-CASCI calculation with CASPT2's NOs seems to be a reasonable and more economic alternative to the DMRG-CASSCF calculation in this case.

\begin{figure}[!htp]
\centering
\includegraphics[scale=0.55]{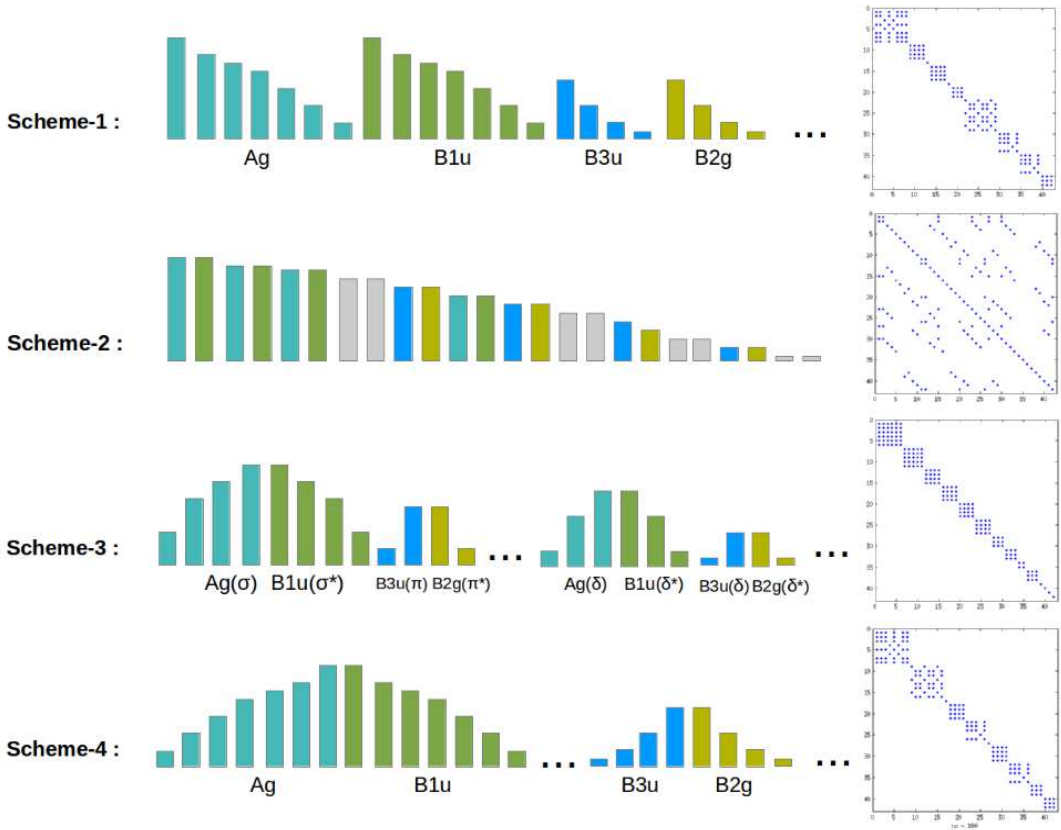}
\caption{Schematic illustration of the four orbital ordering schemes and their approximate entangled range. Here one column represents one orbital and the height of the column reflects the NOONs. The entangled range is measured from one-electron integrals in the Cr$_2$ (r=2.8\AA) system with Space-3.}
\label{fig-schemes}
\end{figure}

From Table.~\ref{tab-Cr2}, we know that the NOs from CASPT2 can provide a good set of initial orbitals at both near-equilibrium and dissociation geometries. Furthermore, it is interesting to see how the results are changed by different orbital ordering schemes using the same orbitals. In the past decade there have been many studies focusing on the orbital ordering in DMRG-based quantum chemical calculations \cite{Chan02, Legeza03, Moritz05}. Here, we make further investigations about the NO ordering effect on the DMRG calculations. Besides the original ordering (Scheme-1) used in Table.~\ref{tab-Cr2}, we also tested three other ordering schemes: a) the orbitals are ordered according to the NOONs (Scheme-2); b) bonding and anti-bonding ordering (Scheme-3); and c) similar with Scheme-1, but inverse orbital ordering in half of the irreducible representation groups (Scheme-4). These schemes are also demonstrated in Fig.~\ref{fig-schemes} and the approximate entangled range in each scheme is also given.

Previous works have observed minimizing the bandwidth of integral matrix may offer an economical method to produce optimal orbital ordering which can minimize the interaction ranges of the Hamiltonian and accordingly minimize the correlation length of the system.\cite{Chan02, Legeza03, Moritz05} The results listed in Table.~\ref{tab-scheme} are generally in agreements with such finding.
From Table.~\ref{tab-scheme}, the bonding and anti-bonding ordering (Scheme-3) provides the best results among the four schemes under the near-equilibrium and dissociation conditions, since in this scheme, the entangled orbitals are close to each other. Scheme-4 is second only to Scheme-3 and the deviations between Scheme-4 and Scheme-3 are less than 1 $mH$ in these tests. Scheme-1, which is also the default ordering scheme in this study, demonstrate a similar precision to Scheme-3 and Scheme-4 under the near-equilibrium condition. However, under the dissociation condition, Scheme-1 provides good result in (12e,28o) DMRG calculation, but this scheme gives relatively poor results in (12e,42o) DMRG calculation, and the deviation is about 8 $mH$ compared to the result of Scheme-3. This unsatisfactory result is caused by the non-optimized entangled range and the insufficient preserved states. The Scheme-2 always gives the worst results, because this ordering will give unbalance "system" and "environment" blocks in the DMRG sweeps and the interactions are mainly localized in the orbitals which own large NOONs, then the lack of the quantum information exchange between blocks will much affect the accuracy \cite{Legeza03}, and this scheme also has the longest entangled range.


\begin{table}[!hbp]
 \centering
 \begin{threeparttable}
 \caption{\label{tab-scheme}Calculated Cr$_2$ ground state correlation energies (a.u.) by DMRG-CASCI ($M$=2000) using CASPT2(12e,12o)'s NOs with various orbital ordering schemes.}
  \begin{tabular}{lcccccccccccccc}
  \hline
  \hline
 &      &&& \multicolumn{3}{c}{1.5\AA}  && \multicolumn{3}{c}{2.8\AA}  \\
 \cline{5-7} \cline{9-11}
 \multicolumn{2}{c}{Orbital ordering} &&&  & (12e,28o) & (12e,42o) &&   & (12e,28o) & (12e,42o)\\
  \hline
 & Scheme-1 &&& & -0.77862 & -0.83859 && & -1.46714 & -1.52019 \\
 & Scheme-2 &&& & -0.77450 & -0.83050 && & -1.45949 & -1.51635 \\
 & Scheme-3 &&& & -0.77879 & -0.83919 && & -1.46718 & -1.52824 \\
 & Scheme-4 &&& & -0.77868 & -0.83843 && & -1.46717 & -1.52737 \\
  \hline
  \hline
  \end{tabular}
 \end{threeparttable}
\end{table}

\begin{figure}[!htp]
\centering
\includegraphics[scale=0.50]{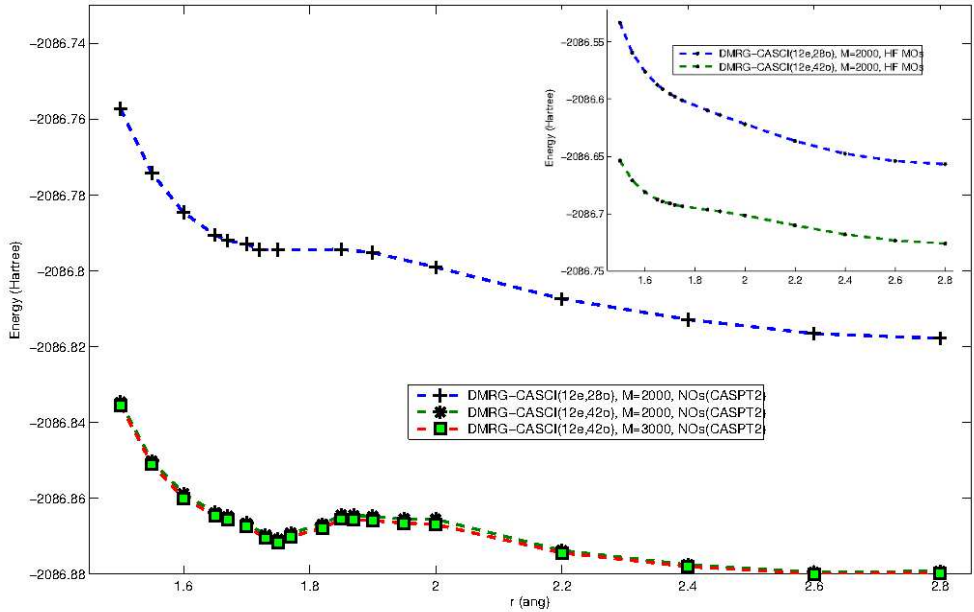}
\caption{Potential curves of Cr$_2$ by DMRG-CASCI calculations using CASPT2's NOs. The cc-pvtz basis and two active spaces ( Space-2(12e,28o) and Space-3(12e,42o) ) are used. The potential curves by DMRG-CASCI calculations using HF MOs are also shown in the upper right.}
\label{fig-Cr2}
\end{figure}

Using the optimized orbital ordering scheme (Scheme-3) and the CASPT2's NOs, we also draw the potential curve of Cr$_2$ by DMRG-CASCI calculation in the Space-2 (12e,28o) and Space-3(12e,42o), as shown in Fig.~\ref{fig-Cr2}. It could be observed that when using the NOs derived from CASPT2, the more reasonable results are obtained compared to those using HF MOs. In Space-2(12e,28o), the potential energy is monotonically decreasing when using HF MOs, while the potential curve exist a platform region from r=1.70\AA \ to r=1.90\AA \ when using CASPT2's NOs. In space-3(12e,42o), a valley emerges near r=1.75 \AA \ in the potential curve when using CASPT2's NOs. This value (1.75\AA ) is still away from the experiment determined equilibrium Cr-Cr bond length (1.679\AA) \cite{Cr2_exper}, however, it implies that with sufficiently large active space and reasonable NOs, the DMRG-CASCI calculations can correctly predict whether a multiple bond can be formed between two transition metal atoms, at least qualitatively. Obviously, further DMRG-CASCI calculations with larger active spaces can present quantitatively more reasonable results.

\subsection{Hydrogen aggregates systems}

\subsubsection{1-D hydrogen chain (H$_{24}$)}

\begin{figure}[!htp]
\centering
\includegraphics[scale=0.6]{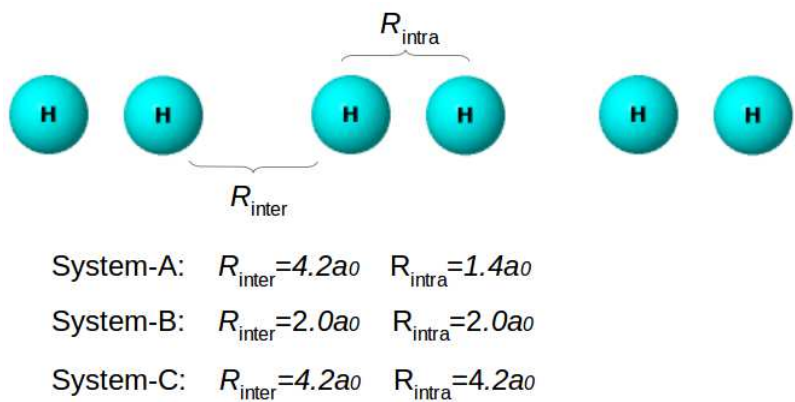}
\caption{The three different geometry configurations used in this subsection}
\label{H_chains}
\end{figure}

Here we consider the 1-D hydrogen chain systems (H$_{24}$). It is well-known that LMOs can provide the optimal basis for DMRG calculations of such 1-D systems. Here, for studying the use of NOs in general systems we only consider such a 1-D system as a general case and accordingly we don't use LMOs. In this system, it is preferable to be treated all electrons as active electrons and a reasonable active space should be at least (24e,24o). Although such a large active space already exceeds the maximum active space of the traditional multi-reference method, the DMRG-CI calculation can still accommodate it. We consider three different geometrical configurations, and we denote them by labels A, B, C, respectively: System-A is dimerized with 1.4 $a_0$ intra-molecular spacing and 4.2 $a_0$ inter-molecular spacing; System-B and System-C have equal intra- and inter-molecular spacings which are 2.0 $a_0$ and 4.2 $a_0$, respectively. These three configurations are also illustrated in Fig.~\ref{H_chains}. Obviously, the polyradical character becomes more and more evident when the system changes from A to B to C.
The basis set used here was Pople's 6-311G basis set, with 72 atomic basis functions. Our preliminary DMRG-CI calculations were performed within the full orbital space (24e,72o), and DMRG-CASCI and DMRG-CASSCF calculations were implemented within two different active spaces (Space-1(24e,24o) and Space-2(24e,36o)) with $M$ = 1000 for these systems.

The calculated ground-state energy results are listed in Table.~\ref{tab-H}. In the 1.4 $a_0$-4.2 $a_0$ case (System-A), the DMRG-CASCI results based on the NOs are similar to not only each other but also to those of DMRG-CASSCF for both Space-1(24e,24o) and Space-2(24e,36o). Typical deviations are about only 2 $mH$. This is because the dynamic correlation for the ground state of this configuration and is the main contribution to the correlation effect, and all of the methods used here in this study can well describe this type of electron correlation, similar to thecase of the N$_2$ and Cr$_2$ systems. Here the result of DMRG-CASCI from CASPT2's NOs is lower than that of DMRG-CASSCF. This is due to the insufficient $M$ values used in DMRG-based calculations utilized in the study. For the uniform spacing case with 2.0 $a_0$ separation (System-B), similar behavior is observed between the results here and those in the 1.4 $a_0$-4.2 $a_0$ case (System-A), since dynamic correlation also plays a most important role here. The DMRG-CASCI results are also similar to those of DMRG-CASSCF. The deviations are less than 4 $mH$ with NOs from MP2, CISD, and CASPT2, and the deviations further reduced (about 1 or 2 $mH$) when using the NOs from DMRG-CI, because DMRG-CI calculations can also give better descriptions for static correlations. For the uniform spacing case with a 4.2 $a_0$ separation (System-C) in which the polyradical and multi-reference characters become significantly stronger, the CASPT2 results could not be obtained become of insufficiently active space, and the DMRG-CASSCF calculations also encountered convergence difficulties. Although the reference DMRG-CASSCF results are not converged, the results based on the above-mentioned NOs still have larger deviations as compared to the DMRG-CASSCF results. In Space-1 (24e,24o), the deviation is about 0.2 hartree for CISD's NOs and is about 0.16 hartree for MP2's NOs, while the deviation is much smaller in the results based on DMRG-CI's NOs (about 0.07 hartree with $M$ = 200 or $M$ = 300). In Space-2(24e,36o), a similar phenomenon is observed. These results implies that, although all NOs obtained by economic electron correlation calculations are not very satisfactory, under conditions in which multi-reference characters are very strong, preliminary DMRG-CI can still provide much more reasonable NOs than other methods. Continue adding the number of preserved states in preliminary DMRG-CI calculations will benefit the quality of NOs, and consequently new efficient DMRG-CI algorithms are greatly desired.

 \begin{table}[!hbp]
 \centering\footnotesize
 \begin{threeparttable}
 \caption{\label{tab-H}Calculated ground state energies (in a.u.) by DMRG-CASCI in the 1-D hydrogen chains systems. The 6-311G basis and D2h symmetry are used. }
 \begin{tabular}{lcccccccccccccccc}
  \hline
  \hline
 &          && \multicolumn{2}{c}{1.4 $a_0$-4.2 $a_0$} && \multicolumn{2}{c}{2.0 $a_0$-2.0 $a_0$} && \multicolumn{2}{c}{4.2 $a_0$-4.2 $a_0$}  \\
  \cline{4-5} \cline{7-8} \cline{10-11}
 & {Orbital type} && (24e,24o)$^a$ & (24e,36o)$^b$ && (24e,24o)$^a$ & (24e,36o)$^b$ && (24e,24o)$^a$ & (24e,36o)$^c$ \\
 \hline
    & DMRG-CI(24e,72o) NOs\\
   & $M$=100 && -13.76357 & -13.81405 && -13.17205 & -13.22172 && -11.86084 & -11.89152 \\
   & $M$=200 && -13.76345 & -13.81399 && -13.17247 & -13.22199 && -11.90353 & -11.89605  \\
   & $M$=300 && -13.76314 & -13.81390 && -13.17258 & -13.22197 && -11.89499 & -11.88834 \\
   \hline
   & NOs (MP2)   && -13.76242 & -13.81365 && -13.16774 & -13.22039  && -11.81066 & -11.78991  \\
   & NOs (CISD)  && -13.76363 & -13.81414 && -13.17100 & -13.22185  && -11.77204 & -11.76828   \\
   & NOs (CASPT2$^d$) && -13.76262 & -13.81625 && -13.16907 & -13.22097  &&     -      &   -      \\
   \hline
   & DMRG-CASSCF && -13.76493 & -13.81431 && -13.17376 & -13.22241 && -11.96922$^{uc}$ & -11.90881$^{uc}$\\
  \hline
  \hline
  \end{tabular}
 \begin{tablenotes}
 \item[a] The active orbitals are 12$A_g$+12$B_{1u}$.
 \item[b] The active orbitals are 18$A_g$+18$B_{1u}$.
 \item[c] The active orbitals are 18$A_g$+18$B_{1u}$ except DMRG-CI NOs. In NOs(DMRG) cases they are 20$Ag$+16$B1u$.
 \item[d] 8e,8o form the active space in CASPT2 calculation.
 \item[uc] This results is not converged.
 \end{tablenotes}
 \end{threeparttable}
\end{table}

\subsubsection{2-D Hydrogen system (H$_{16}$ and H$_{24}$)}

\begin{figure}[!htp]
\centering
\includegraphics[scale=0.5]{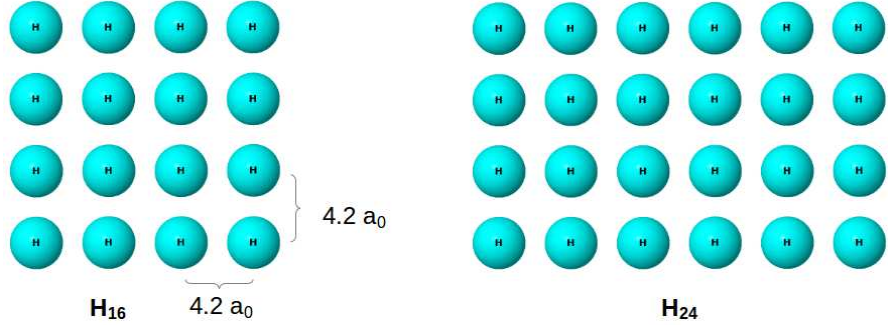}
\caption{Geometries of H$_{16}$ and H$_{24}$ systems}
\label{H_2D}
\end{figure}

Finally, we consider the more general 2-D Hydrogen systems (H$_{16}$ and H$_{24}$). In these two systems, one can use two additional irreducible representation (totally $A_g$, $B_{2u}$, $B_{1u}$, $B_{3g}$) than in 1-D systems ($A_g$, $B_{1u}$), and as such, the DMRG-CASCI calculations can be implemented with much more preserved states than in 1-D case. The intermolecular distance is 4.2 $a_0$, and the geometries are shown in Fig.~\ref{H_2D}. At this intermolecular distance, the system will posses strong polyradical and multi-reference characters, as was demonstrated for the 1-D hydrogen chain.

Here Pople's 6-311G basis sets were used, and the total number of orbitals were 48 (H$_{16}$) and 72 (H$_{24}$). The NOs from CISD and MP2 calculations were chosen, and the NOs from the MRCI calculation were used instead of those from CASPT2. Preliminary DMRG-CI calculations were performed within the full orbital space with gradually increasing $M$ values. The DMRG-CASCI and DMRG-CASSCF calculations were implemented within their valence active spaces with $M$ = 6000 (H$_{16}$) and $M$ = 4000 (H$_{24}$). The FCI calculations are also implemented by Molpro in H$_{16}$ and taken as references. The results are shown in Table.~\ref{tab-2DH}. In the H$_{16}$ system, DMRG-CASCI from preliminary DMRG-CI's NOs ($M$ $\leq$ 500) provides good results, and the deviation is less than 8 $mH$ compared to the DMRG-CASSCF value. The NOs from single-reference methods such as MP2 and CISD give poor results in contrast to the CASSCF or DMRG-CASSCF results because of the lack of the static electronic correlation. The NOs from MRCI give much better results than those from MP2 and CISD, and the deviation is about 28 $mH$ compared to the DMRG-CASSCF result. This is because the CASSCF(8e,8o) are used as the reference state in the MRCI calculation and as such part of the important static correlations can be recovered.
For the H$_{24}$ system, the NOs from CISD are still not suitable, and the NOs from MRCI show no obvious advantage over those of CISD, even not as good as the NOs derived from MP2 calculations, which is due to the fact that the CASSCF(8e,8o) is insufficient to recover the important static correlations in this system. Preliminary DMRG-CI is somewhat inefficient owing to the large number of total orbitals, but DMRG-CI(24e,72o) calculation with $M$ = 1000 still provides relatively most reasonable NO basis, leading to a deviation of about 30 $mH$ compared to DMRG-CASSCF result. Such deviation is less than those of all other results.

\begin{table}[!hbp]
 \centering
 \begin{threeparttable}
 \caption{\label{tab-2DH}Calculated ground state energies (in a.u.) by DMRG-CASCI in the 2-D hydrogen aggregate systems. The 6-311G basis and D$_{2h}$ symmetry are used.}
 \begin{tabular}{lcccccccccccccccc}
  \hline
  \hline
 &          & \multicolumn{2}{c}{H$_{16}$} && \multicolumn{1}{c}{H$_{24}$}   \\
  \cline{3-4} \cline{6-6}
 & {Orbital type} &  M6000$^a$ & FCI(16e,16o)$^a$ && M4000$^b$ \\
 \hline
    & NOs from DMRG-CI \\
   & $M$=100 & -8.00178 &    -     &&  -11.93219 \\
   & $M$=500 & -8.05351 &    -     &&  -11.94893 \\
   & $M$=1000& -8.05331 &    -     &&  -11.98926  \\
   \hline
   & NOs (MP2)   & -8.01091 & -8.01811 && -11.94029  \\
   & NOs (CISD)  & -7.99380 & -8.00037 && -11.84704  \\
   & NOs (MRCI)$^c$  & -8.03345 & -8.04079 && -11.91458  \\
   \hline
 & CASSCF/DMRG-CASSCF & -8.06135  & -8.06921 && -12.02066$^d$  \\
  \hline
  \hline
  \end{tabular}
 \begin{tablenotes}
 \item[a] The active orbitals are 4$A_g$+4$B_{2u}$+4$B_{1u}$+4$B_{3g}$.
 \item[b] The active orbitals are 6$A_g$+6$B_{2u}$+6$B_{1u}$+6$B_{3g}$.
 \item[c] 8e,8o form the active space in MRCI calculation.
 \item[d] This value is estimated as the sum of the ground sate energy of DMRG-CASCI ($M$=4000) and the ground state energy difference between DMRG-CASCI and DMRG-CASSCF ($M$=2000), since the cost of DMRG-CASSCF ($M$=4000) is expensive.
 \end{tablenotes}
 \end{threeparttable}
\end{table}

\clearpage
\section{Summary and conclusion}

In this study, we investigated the performances of various NOs and HF CMOs, KS orbitals, as well as CASSCF-optimized orbitals, as the basis sets in \emph{ab initio} DMRG-CASCI calculations with large active space and compare them with DMRG-CASSCF calculations.

We observed that CASPT2 and MRCI NOs always gives reasonable basis sets for further higher-level DMRG-CASCI calculations provided that a reasonable active space for the CASPT2 and MRCI calculations. The deviations of DMRG-CASCI energies using CASPT2 NOs from DMRG-CASSCF results are less than 4 $mH$ in our N$_2$ and Cr$_2$ systems under both near-equilibrium and dissociation conditions.
Simultaneously, single-reference CISD NOs also provide good basis sets for further higher-level DMRG-CASCI calculations utilizing sufficiently large active spaces, because such DMRG-CASCI calculations can recover static correlations that are poorly described by CISD.

The NOs generated from the DMRG-CI calculations are invariably reasonable, provided the number of preserved states $M$ is sufficiently large. When the multi-reference character is not strong, a small number of preserved states (eg. $M$=100) usually provide a good set of NOs for our test systems; while under the bond dissociation situation, much more preserved states are needed to recover the static correlations. For example in the N$_2$ system, only $M$ = 500 leads to a result better than derived that from CASPT2 NOs. However, it is not necessary to use DMRG-CI to generate NOs for high-level DMRG-CASCI calculations in most cases, since the computational costs of the DMRG algorithm usually exceeds those of traditional electron correlation methods, such as MP2 and CISD as well as the CASPT2.
Here we should also mention that the NOs from DMRG-CI may lose orbital degeneracy because of the insufficiently preserved states and also the ``sweep" process in DMRG calculations. If necessary, this can be corrected by the DMRG-CASSCF procedure with a few macro-iterations (typically 1$\sim$2, since the orbitals are already optimized in the moderate DMRG-CI level).

However, it is worth noting that a priori DMRG-CI calculations are mandatory when the multi-reference character is very strong (for example, in our 1-D and 2-D hydrogen polyradical system). In such case, describing static correlations requires an active space composed of more than 20 orbitals, which is far beyond the capability of current CASSCF or CASPT2 method. Only the DMRG-CI can provide significantly more reasonable natural orbitals than MP2, CISD, CASPT2, and MRCI.

Our results reveal that a good compromise between the requirement for low computational costs of acquiring NOs and the demand for high efficiency of NOs as the basis of DMRG calculations may be differ according to the studied systems' diverse electron correlation characteristics and the size of the active space. For DMRG calculations with medium-sized active spaces, quantum chemical methods like CASPT2 and MRCI which can well account for static and dynamic electron correlations, are necessary to generate the NO basis set. When the active space becomes sufficiently large, quantum chemical methods such as CISD, which can efficiently describe dynamic electron correlations, are sufficient to produce reasonable NO basis for DMRG calculations.
As for cases with very strong multi-reference characters that involves more than 17 unpaired electrons, preliminary DMRG-CI calculations are preferable for obtaining more reasonable NOs for further higher-level DMRG-based calculations.
Such findings may provide useful guidance for the selection of suitable quantum chemical method to generate reasonable NOs as the basis for subsequent large-scale DMRG-based calculations of different types of systems.

Our results also show that DMRG-CASCI calculations from a basis of suitable NOs can obtain high accuracies, which are comparable to those of DMRG-CASSCF calculation. This strategy can overcome the disadvantage of traditional DMRG-CASSCF (i.e., the convergence difficulty) and reduce the computational costs in ``macro-iterations". Actually, the ``three stages" DMRG-SCF strategy proposed by Zgid and Nooijen \cite{dmrg-scf}, in which DMRG calculations are performed with a gradually increasing small \emph{M} value during orbital optimization, can also remarkably decrease the computational load compared to the standard DMRG-CASSCF calculations. However, the two preconditions required for this ``three stages" DMRG-SCF strategy, no convergence problem and inexpensive integral transformation, can not always be fulfilled for complicated systems with very large active spaces.

\section*{Acknowledgment}
This work is supported by the National Basic Research Program (Grant No. 2011CB808604), National Natural Science Foundation of China (Grant No. 21003072 and 91122019),  and the Fundamental Research Funds for the Central Universities. We thank C. Liu and W. Hu for the stimulating discussions about orbital optimization algorithm in DMRG-CASSCF calculations.

\section*{Appendix: Chosen orbitals in Table.~\ref{tab-Cr2}}

 \begin{table}[!hbp]
 \begin{threeparttable}

 \begin{tablenotes}
 \item[a] Frozen orbitals are (5,2,2,0,5,2,2,0) with irreducible representation order $A_g$, $B_{3u}$, $B_{2u}$, $B_{1g}$, $B_{1u}$, $B_{2g}$, $B_{3g}$, $A_{u}$ in all cases.
 \item[b] Active orbitals are (3,1,1,1,3,1,1,1).
 \item[c] Active orbitals are (7,4,4,2,6,2,1,2) in NOs($M$=100), (7,4,4,3,5,2,1,2) in NOs($M$=200) and (7,4,3,3,5,2,2,2) in NOs($M$=300).
 \item[d] Active orbitals are (6,3,3,2,6,3,3,2) in HF, KS and CASSCF MOs, (7,5,5,3,3,2,2,1) in NOs(CISD) and (8,5,5,3,4,1,1,1) in NOs(CASPT2).
 \item[e] Active orbitals are (11,6,6,3,6,3,4,3) in NOs($M$=100), (10,6,6,3,7,4,4,2) in others.
 \item[f] Active orbitals are (8,5,5,3,8,5,5,3) in HF, KS and CASSCF MOs, (9,6,6,3,7,4,4,3) in others.
 \item[g] Active orbitals are (8,3,3,2,6,2,2,2) in NOs($M$=100) and NOs($M$=300), (8,3,3,2,5,3,2,2) in NOs($M$=200).
 \item[h] Active orbitals are (6,3,3,2,6,3,3,2).
 \item[i] Active orbitals are (11,5,5,2,10,3,4,2) in NOs($M$=100), (10,5,5,2,10,4,4,2) in others.
 \item[j] Active orbitals are (8,5,5,3,8,5,5,3).

 \item[k] Active orbitals are (7,4,4,3,4,2,2,2).
 \item[l] Active orbitals are (9,6,6,3,7,4,4,3).
 \item[m] Active orbitals are (8,3,3,2,6,2,2,2).
 \item[n] Active orbitals are (10,6,5,2,8,5,4,2).
 \item[o] 12e,12o form the active space.
  \end{tablenotes}
  \end{threeparttable}
\end{table}

\end{document}